\documentstyle[12pt]{article}

\begin{document}
\title{Study of Spiral Pattern Formation in Rayleigh-B\'enard Convection }
\author{Hao-wen Xi\\
Department of Physics\\
Lehigh University\\
Bethlehem, Pennsylvania 18015,\\
J.D. Gunton\\
Department of Physics\\
Lehigh University\\
Bethlehem, Pennsylvania 18015,\\
and\\
Jorge Vi\~nals\\
Supercomputer Computations Research
Institute, B-186\\
and\\
Department of Chemical Engineering, B-203\\
Florida State University\\
Tallahassee, Florida 32306-4052.}
\maketitle
\newpage

We present a numerical study of a generalized two-dimensional
Swift-Hohenberg model of spiral pattern formation in
Rayleigh-B\'enard convection in a non-Boussinesq fluid.
We demonstrate for the first time that a model for convective
motion is able to predict in considerable dynamical detail the spontaneous
formation of a rotating spiral state from an ordered hexagon
state. Our results are in good agreement with recent experimental studies of
$CO_{2}$ gas. The mean flow and non-Boussinesq effects
are shown to be crucial in forming rotating spirals.

\newpage

One of the most striking examples of spatio-temporal self-organized
phenomena in non-equilibrium systems is the rotating spiral seen in
chemical and biological systems \cite{re:ni77} \cite{re:cb91}.
The Belousov-Zhabotinsky (BZ) reaction \cite{re:bz91} has received
considerable attention as an example of chemical wave propagation.
The formation of spiral patterns in the BZ system resulting from the
coupling of a reaction process with a transport process such
as diffusion has been extensively studied, both theoretically and
experimentally, during the past
decade. Remarkably, similar rotating spirals were observed recently in
Rayleigh-B\'enard (RB) convection in non-Oberbeck-Boussinesq
fluids \cite{re:bo91}. Much of the earlier experimental work has been
restricted to Oberbeck-Boussinesq type fluids, in which one observes various
configurations of roll patterns. However, in a non-Oberbeck-Boussinesq
system with, for example, temperature dependent transport coefficients,
both roll and hexagonal patterns can exist. Very recently
Bodenschatz et al \cite{re:bo91} have performed experiments on convection
in $CO_{2}$ gas and studied the existence of and transitions
between convective patterns exhibiting different symmetries.
They have observed the competition between a uniform conducting
state, a convective state with hexagonal symmetry, and
a convective state consisting of rolls. Their most surprising discovery
is that the hexagon-roll transition has a tendency to form rotating spirals.
In this paper we present the first numerical evidence for the spontaneous
formation of a rotating spiral pattern during the hexagon-roll transition
in a large aspect ratio cylindrical cell near onset. We also report the
numerical results of spiral formation during the conduction-roll transition.
We will qualitatively compare our numerical results with the experiments of
Bodenschatz et al.

We now describle our two-dimensional (2D) study of spiral patterns formed
in a thin layer of a non-Oberbeck-Boussinesq fluid. We model such a fluid
by a two dimensional generalized Swift-Hohenberg equation \cite{re:sw77}
\cite{re:ma83}, given by equations (1) and (2) below, which we solve
by numerical integration. The Swift-Hohenberg equation and various
generalizations of it have proven to be quite successful in explaining
many features of convective flow \cite{re:gr82},
particularly near onset. As we show
in this paper, the same holds true for the non-Oberbeck-Boussinesq
fluid. Our model is defined by
\begin{equation}
\label{eq:sh}
\frac{\partial \psi (\vec{r},t)}{\partial t} +g_{m} \vec{U} \cdot \nabla \psi=
\left[ \epsilon - \left( \nabla^{2} + 1 \right)^{2} \right] \psi
-g_{2} \psi^{2} - \psi^{3} + f(\vec{r}),
\end{equation}

\begin{equation}
\label{eq:mean}
\left[ \frac{\partial }{\partial t} -Pr( \nabla^{2}-c^{2} ) \right]
\nabla^{2} \xi = \left[ \nabla \psi \times \nabla(\nabla^{2} \psi) \right]
\cdot \hat{e}_{z},
\end{equation}

\noindent where,
\begin{equation}
\vec{U}=(\partial_{y} \xi) \hat{e}_{x} - (\partial_{x} \xi) \hat{e}_{y}
\end{equation}

\noindent with boundary conditions,
\begin{equation}
\psi|_{B} = \hat{n} \cdot \nabla \psi |_{B} = 0,
\end{equation}
where $\hat{n}$ is the unit normal to the boundary of the domain of
integration, $B$.
This equation with $g_{2}=g_{m}=0$ reduces to the Swift-Hohenberg
(SH) equation and
has been extensively used to model convection in thin cells and near onset
\cite{re:gr82,re:co90,re:ma90,re:cr80}. The scalar order parameter
$\psi(\vec{r},t)$ is related to the fluid temperature in the
mid-plane of the convective cell. $\xi(\vec{r},t)$ is the vertical vorticity
potential. This mean flow field coupling \cite{re:ma83,re:si81}
with the
SH equation has been shown to play a key role in the onset of turbulence
in Oberbeck-Boussinesq systems \cite{re:gr88}. The quantity $\epsilon$ is
the reduced Rayleigh number,
\begin{equation}
\epsilon = \frac{R}{R_{c}^{\infty}} -1,
\end{equation}
where $R$ is the Rayleigh number and $R_{c}^{\infty}$ is the critical Rayleigh
number for an infinite system. A phenomenological forcing field $f$ has been
included in Eq.(1) to simulate the lateral sidewall forcing produced by
horizontal temperature gradients present in the experiment. As in earlier
studies \cite{re:xi91,re:vi91}, we have varied the strength and spatial
extent of $f$ in order to best fit the experimental observations.
We have derived a three mode amplitude equation from the generalized SH
equation in order to both estimate the threshold values of $\epsilon$
that separate regions in which roll and hexagonal configurations are stable,
as well as the values of the parameters that enter the generalized SH equation
in terms of experimentally measurable quantities. From the experiment
\cite{re:bo91}, we find that $g_{2} \approx 0.35$, which is the value that we
have used in Eq.(1).  The value of $\epsilon$ used in the
numerical simulation is related to the real experimental value $\epsilon^{exp}$
in ref. \cite{re:bo91} by $\epsilon^{exp}=0.3594 \epsilon$.
The nonlinear coupling constant $g_{m}$ has been chosen to be 50,
which is consistent with earlier studies \cite{re:si81,re:ma83}, and
$c^{2}=10$ throughout our calculations.
We note that Bestehorn et al \cite{re:be92}
have reported a study of this model limited to the special case in which the
initial configuration is a spiral. This, however, avoids the fundamental
question addressed here of how the spiral spontaneously forms in the
hexagon-roll transition. In the following we report the
results of our calculations.

i) Formation of hexagonal pattern with sidewall forcing field.
We have considered a circular cell of radius $R = 32 \pi$, which
corresponds to an aspect ratio of $\Gamma =2R / \pi = 64$. A square grid with
$N^{2}$ nodes has been used with spacing $\Delta x = \Delta y = 64
\pi$/N, and $N=256$. We approximate the boundary conditions on $\psi$ by
taking $\psi (\vec{r},t) = 0$ for $\| \vec{r} \| \ge R$, where $\vec{r}$
is the location of a node with respect to the center of the domain of
integration. The initial condition $\psi (\vec{r},t=0)$ is a random
variable, gaussianly distributed with zero mean and a variance of $10^{-1}$.
In this case $\epsilon = 0.1$, and $f=0.0$ everywhere except on the nodes
adjacent to the boundary with $f = 0.1$. Figure 1 presents a typical
configuration for a hexagonal pattern which forms in
the presence of a strong static sidewall forcing.

ii) Early stage of transition between hexagons and rolls.
We use Fig. 1 as the initial configuration with exactly the same forcing field
$f$ as before, but now we increase $\epsilon$ very slowly up to 0.3. We take
$\epsilon=0.1+1.67\times 10^{-4}t$ for $0<t<1200$ and $\epsilon=0.3$
for $t>1200$. Figure 2 shows two configurations
during the early transient regime during the hexagon to roll
transition. How the sidewall forcing
and defects mediate the transition can be clearly seen in Fig. 2. The rolls
are formed near the sidewall with a favorable orientation relative to the
symmetry of the sidewall. In the meantime the defects glide toward each
other and invade nearby regions of hexagonal order to create a region
of rolls that spreads across the cell as the transition proceeds.
The spiral formation is already noticeable in Fig. 2(b). This resembles
the experimental observation that there is a tendency to form
spirals during the hexagon to roll transition.

iii) Formation of rotating spirals.
Fig. 3 show the spatial and temporal formation of the spiral pattern
at later times than in Fig. 2. In Fig. 3(a), we see that rolls bend or curve
rapidly forming a roughly uniform patch of rolls with a locally disordered
texture near the left corner.  Further evolution consists primarily of
dislocations gliding toward each other and eventually annihilating themselves,
ending in a three-armed spiral. The final state of the rotating spiral
(Figs. 3(b) and 3(c)) is remarkably similar to one observed in the
experiment and occurs at $t \approx 49000 \approx 12$ horizontal-diffusion
times \cite{re:note92}. The corresponding experimental times are in the
range of 10 to 20 horizontal-diffusion times.
Our numerical investigation indicates that the non-Boussinesq effect plays a
crucial role in forming spontaneous spirals. In the absence of $g_{2}$ (with
or without the mean flow field), if we start with a random
initial condition, there is no occurence of a spiral pattern. This strongly
suggests that the formation of spiral patterns is an intrinsic property
of non-Boussinesq systems.
We have studied the transition from a conduction state to a rotating
spiral state, as well as the transition from a rotating spiral state to a
hexagon state. We have also studied the effect of decreasing $\epsilon$
in the hexagon to rotating spiral transition. We find that
a stable n-armed spiral tends toward
one with fewer arms when $\epsilon$ is decreased, in agreement with
experimental observation.
The details of all the above will be discussed elsewhere.

In summary, we have investigated the question of pattern formation in
a model of convection in a non-Boussinesq fluid that allows
patterns of various symmetries. We start with a random initial condition
and show that this leads to the ordered hexagonal
state observed in the experiment. We then show that upon increasing $\epsilon$
we see a dynamical evolution to a new roll state
which contains a rotating spiral pattern. These results are in very good
agreement with the experimental studies in $CO_{2}$ gas.

We wish to thank E. Bodenschatz, G. Ahlers
and D. Cannell for suggesting the numerical investigation of the generalized
Swift-Hohenberg equation, and them and P.C.Hohenberg for many stimulating
conversations and comments. We would like to acknowledge the
assistance given to us by G.Golub and P.E.Bj$\phi$rstad
during the development of the numerical code.
This work was supported in part by the National Science Foundation under
Grant No. DMR-9100245. This work is also supported in part by the
Supercomputer Computations Research Institute, which is partially funded
by the U.S. Department of Energy
contract No. DE-FC05-85ER25000. The calculations reported here have been
carried out on the Cray Y-MP at the Pittsburgh Supercomputing Center.

\newpage

{\bf Figure captions}

Figure 1. Hexagonal pattern starting from random initial condition
obtained in a cylindrical cell with aspect ratio $\Gamma=64$. The values of
the parameters used are $g_{2} = 0.35$, $g_{m}=50$ and $\epsilon = 0.1$. A
non-zero forcing field localized at the boundary with $f = 0.1$ has
been used.

Figure 2. We observe an early stage of hexagon-roll transition obtained
by changing $\epsilon$ from $\epsilon=0.1$ to $\epsilon=0.3$, in a cylindrical
cell with an aspect ratio $\Gamma=64$. The initial condition is the
uniform hexagonal pattern shown in Fig. 2. Two different times,
$t=720$ (b) and $t=960$ (b) are shown.  The rolls appearing near
the defects and sidewall boundaries spreads through the cell as
the transition proceeds.

Figure 3. Pattern formation of a rotating three-armed spiral obtained
in a cylindrical cell with aspect ratio $\Gamma=64$, with $g_{2} = 0.35$,
$g_{m}=50$, $\epsilon = 0.3$ and f=0.1. Time series of the pattern evolution
at (a) t= 7440, (b) t=48240 and (c) t=64080. The final rotating spiral pattern
is shown in (b) and (c).


\begin{thebibliography}{99}
\bibitem{re:note92} We note that the vertical-diffusion time is
$t_{h}/ \Gamma^{2}$.
\end{thebibliography}
\end{document}